\documentstyle[twoside,fleqn,espcrc2,epsf]{article}

\newcommand\csw{c_{\rm SW}}
\newcommand\psibar{\overline\psi}
\newcommand\Dslash{{\smash{\raise 0.16ex \hbox to 0pt{\kern 0.22em $/$
\hss}}D}}
\newcommand{\MSbar}{{\overline{MS}}}
\newcommand{\order}{{\mathcal{O}}}

\title{%
\vspace{-3.1cm}
\begin{flushleft}
       {\normalsize DESY--97-198} \\[-0.2cm]
       {\normalsize HUB--EP--97/54}  \\[-0.2cm]
       {\normalsize September 1997}   \\
\end{flushleft}
       \vspace{0.7cm}
New results for non-perturbative $\order(a)$ improvement in light hadrons
\thanks{Talk presented by P.W. Stephenson}}

\author{M.~G\"ockeler%
           \address{Institut f\"ur Theoretische Physik, Universit\"at
             Regensburg, D-93040 Regensburg, Germany},
        R.~Horsley%
           \address{Institut f\"ur Physik, Humboldt-Universit\"at zu Berlin,
                    Invalidenstra{\ss}e 110, D-10115 Berlin, Germany},
        H.~Perlt%
           \address{Inst.\ f.\ Theo.\ Phys., Universit\"at Leipzig,
                    Augustusplatz 10--11, D-04109 Leipzig, Germany},
        P.~Rakow$^{\rm c}$,
        G.~Schierholz$^{\rm c,\hspace{-0.1cm}}$
           \address{Deutsches Elektronen-Synchrotron DESY,
                    Notkestra{\ss}e 85, D-22603 Hamburg, Germany},
        A.~Schiller$^{\rm d}$
            and
        P.~Stephenson$^{\rm c}$}

\begin{document}
\begin{abstract}
  We have results from light hadron simulations in quenched QCD at
  $\beta=6.0$ and $6.2$ using non-perturbatively improved
  Sheikholeslami--Wohlert fermions in an effort to remove all
  $\order(a)$ effects.  From looking at hadron masses and splittings
  and the RG-invariant quark masses (where we have one point at
  $\beta=5.7$) we suggest this is plausible even
  with the limited data set. The interpretation of the decay constants
  appears to be less clear.
\end{abstract}

\maketitle

\section{INTRODUCTION}

As part of the QCD Structure Function project, we have been looking at
the effect of non-perturbative improvement~\cite{fromalpha} of Wilson
fermions with the Sheikholeslami--Wohlert (SW) term in quenched QCD.
In this contribution we describe our principal results.  A more
detailed description has recently appeared in~\cite{ournewpaper}.

The method is now standard.  We use the SW term,
\begin{equation}
S_{\rm SW} = \frac{i}{2} \kappa g \csw a
  \sum_{x}\psibar(x)\sigma_{\mu\nu}F_{\mu\nu}(x)\psi(x) \label{eq:sw}
\end{equation}
with the coefficients~\cite{fromalpha} $\csw = 1.769$ at $\beta = 6.0$ and
$\csw = 1.614$ at $\beta = 6.2$.  All the data shown here either uses
these values or is unimproved Wilson data.

\begin{figure}[thb]
\vspace*{-0.2cm}
\hspace*{0.3cm}
\epsfxsize=6.5cm \epsfbox{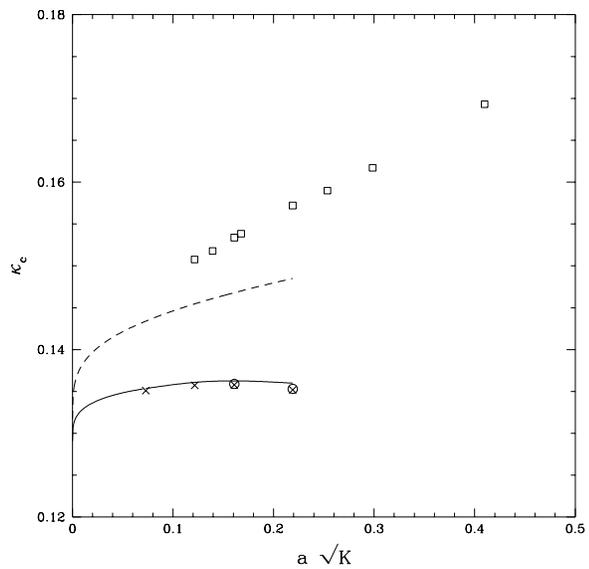}
\vspace*{-1.80cm}
\caption{\footnotesize Monte Carlo and tadpole-improved perturbation
  theory values for $\kappa_c$.  The dashed line and squares are for
  Wilson fermions, the solid line, crosses (our data) and circles
  (ALPHA collaboration data) for improved fermions.}
\vspace*{-0.8cm}
\label{showingkappac}
\end{figure}

Apart from improvement of the
action, matrix elements and renormalisation require an additional
improvement to remove $\order(a)$ effects.
Ideally these should
be calculated non-perturbatively, but for the time being some of our
coefficients have come from tadpole-improved perturbation theory.
However, the effectiveness of this procedure is much greater than in
the Wilson case, as can be seen in figure~\ref{showingkappac} which
compares the predictions for the critical hopping parameter with the
values from the Monte Carlo data.

\section{LIGHT HADRON MASSES}

\begin{figure}[hbt]
\vspace*{-0.5cm}
\hspace*{0.35cm}
\epsfxsize=6.5cm \epsfbox{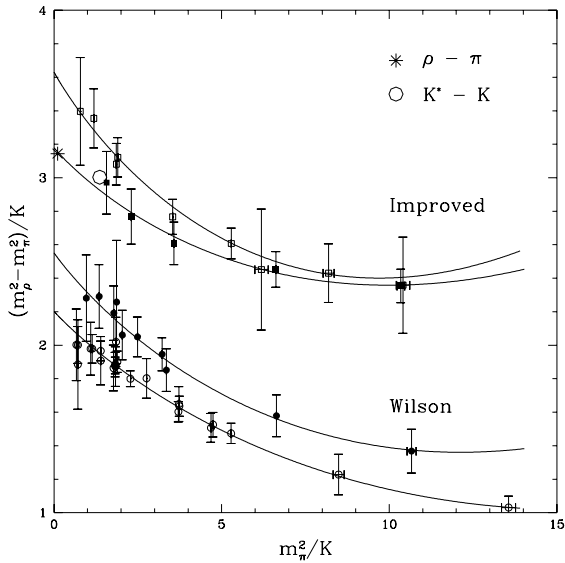}
\vspace*{-1.90cm}
\caption{\footnotesize The vector--pseudoscalar mass splitting for
  Wilson (circles) and improved (squares) fermions.  Open symbols are
  $\beta=6.0$ and filled symbols $\beta=6.2$.}
\vspace*{-0.5cm}
\label{withsplittings}
\end{figure}

Our Edinburgh/APE plots confirm that there is a better behaviour as
one approaches the chiral limit, however the errors here are too large
to draw strong conclusions.  New results for large lattices at
$\beta=6.2$ are in production~\cite{byDirk}.  A useful test of
improvement is in the splitting of the pseudoscalar and vector masses.
This is not well described by Wilson data.  In
figure~\ref{withsplittings}, we show the difference in the squares of
the vector and pseudoscalar masses for both Wilson and improved data
at both beta values against the pseudoscalar mass.  The improved data
is now consistent with the physical values for both light quark
(pion/rho) and strange quark ($K$/$K^*$) masses.  The string tension
has been used to set the scale.

Another parameter which has been used is
the J parameter~\cite{duetolacockandmichael}
which investigates the slope of the graph (assumed linear) rather than
the absolute
value.  Here we see instead no significant improvement towards the
physical value of $0.49$:  at $\beta=6.0$, in fact, the value has
changed with improvement from $0.413(6)$ to $0.38(2)$, and at $6.2$
there is only an insignificant change in the other direction, from
$0.40(3)$ to $0.42(3)$.  It remains to be seen whether this
discrepancy comes from differing discretisation errors in the
dependence of the vector and pseudoscalar masses on the quark mass
(and can therefore be rescued by performing separate extrapolations to
the continuum limit at different quark masses), or whether it is a
more fundamental problem with the quenched approximation.  Our results in
figure~\ref{withsplittings} provide scant support for suggestions of any
intrinsic problems with the masses themselves, even in the chiral limit.

\section{QUARK MASSES}

We have calculated the light and strange quark masses using two
substantially different methods which allows us to check the
consistency in the continuum limit. 

The first method is the traditional one:  the masses are deduced from
the bare values contained in $\kappa$.  A single overall
renormalisation factor $Z_m$ is required; this is scale dependent and
we pick the common value of 2 GeV as the scale in the $\MSbar$ scheme.
In the second method, we use the PCAC relation to deduce the sum of
the light quark masses from the axial and pseudoscalar currents $A$
and $P$.  Here $Z_P$ carries the scale dependence.

\begin{figure}[hbt]
\vspace*{2.1cm}
\hspace*{-1.5cm}
\epsfxsize=9cm \epsfbox{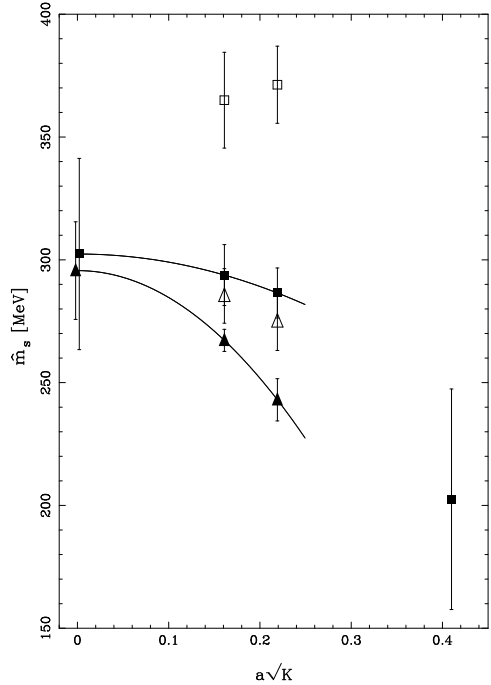}
\vspace*{-0.7cm}
\caption{\footnotesize The strange quark mass from the standard method
  (squares) and the PCAC Ward identity (triangles) for Wilson (open) and
  improved (filled) quarks.}
\vspace*{-0.85cm}
\label{showingstrangequarkmass}
\end{figure}

In figure~\ref{showingstrangequarkmass} we show the strange quark
mass.  In addition to our $\beta=6.0$ and $6.2$ results, we show a
single point for the standard method at $\beta=5.7$ using an
improvement coefficient $\csw=2.25$, which is near to and probably
slightly above the value required for full $\order(a)$
improvement~\cite{atleastTimsaysso}.  We have again relied on tadpole
improved perturbation theory for various coefficients.

The quantity displayed is the renormalisation group invariant mass
$\hat m$;
our definition of this is~\cite{whichisfromhere} (see reference for
values of quantities):
\begin{equation}
m(\mu^2) = \hat m\left(\frac{\alpha_s(\mu^2)}{\pi}\right)^{\gamma_0/2\beta_0}
(1 + A_1\frac{\alpha_s(\mu^2)}{\pi} + \cdots)
\end{equation} 
The formula has been used to two loops to produce the results shown.
Dividing the invariant masses shown in the graph by 1.65 roughly gives
the values normalised at 2.0 GeV.

We use a quadratic extrapolation with no linear term for the improved
data, not including the $\beta=5.7$ result, so the lines are exactly
determined.  It is clear that the agreement is much better in the
improved case, largely due to the movement of the result from the
traditional method into line with the others.  Maybe this is
connected with the improved $\kappa_c$ behaviour noted above.

\section{DECAY CONSTANTS}

\begin{figure}[hbt]
\vspace*{-0.4cm}
\hspace*{0.3cm}
\epsfxsize=6.5cm \epsfbox{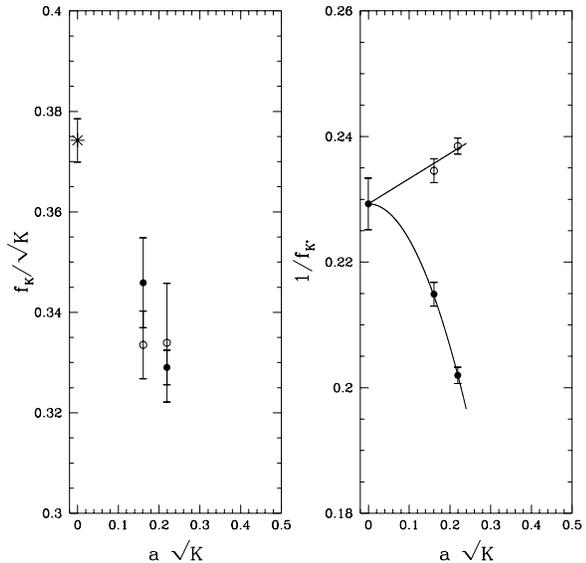}
\vspace*{-1.90cm}
\caption{\footnotesize $K$ and $K^*$ decay constants,
 with Wilson (open) and
   improved (filled) fermions.  The experimental value (star) is
   shown for the $f_K$.}
\vspace*{-0.6cm}
\label{theonewithKandKstar}
\end{figure}

Our decay constants come from observables with a smeared source and a
local sink.  In the case of the $f_\pi$ and $f_\rho$ it is less clear
that one can validly claim $\order(a)$ errors in the one case and
$\order(a^2)$ in the other.  We show the values at the strange quark
mass for the $K$ and $K^*$ in figure~\ref{theonewithKandKstar}.  The
discretisation errors are larger with the improved action and it is
difficult to come to further conclusions about the scaling behaviour,
which seems here to be worse than in the chiral limit.  We note that
the contribution of the improvement terms is also larger in this
quark mass region.

\section{SUMMARY}

Although we have only two values of the coupling, our data from light
hadron and quark masses appears to be consistent with the removal of
all $\order(a)$ effects by the non-perturbatively improved fermion
action.  The splitting of the vector and pseudoscalar masses is now in
agreement with experiment comfortably within our errors.  New
larger lattice data at $\beta=6.2$~\cite{byDirk} will extend this work.
This work was supported in part by the Deutsche Forschungsgemeinschaft.

\end{document}